\documentclass[conference]{IEEEtran}
\usepackage{csquotes}
\usepackage[section]{placeins}

\usepackage{cite}
\ifCLASSINFOpdf
   \usepackage[pdftex]{graphicx}
   \graphicspath{{../pdf/}{../jpeg/}}
   \DeclareGraphicsExtensions{.pdf,.jpeg,.png}
\else
   \usepackage[dvips]{graphicx}
   \graphicspath{{../eps/}}
   \DeclareGraphicsExtensions{.eps}
\fi
\usepackage{multirow}
\usepackage[table]{xcolor}
\usepackage{booktabs}
\ifCLASSOPTIONcompsoc
  \usepackage[caption=false,font=normalsize,labelfont=sf,textfont=sf]{subfig}
\else
  \usepackage[caption=false,font=footnotesize]{subfig}
\fi
\usepackage{float}

\usepackage[acronym, nowarn]{glossaries}
\makeglossaries
\newacronym{manet}{\textsc{MANET}}{mobile ad hoc networks}
\newacronym{tetra}{\textsc{TETRA}}{Terrestrial Trunked Radio}
\newacronym{hf}{\textsc{HF}}{High Frequency}
\newacronym{uhf}{\textsc{UHF}}{Ultra High Frequency}
\newacronym{vhf}{\textsc{VHF}}{Very High Frequency}
\newacronym{sbd}{\textsc{SBD}}{short-burst-data}

\usepackage[binary-units=true]{siunitx}
\usepackage{hyperref}
\usepackage[hyphenbreaks]{breakurl}

\usepackage{tikz}
\usetikzlibrary{calc, positioning, arrows, shapes, shadows, trees, arrows.meta, 
decorations.pathreplacing, hobby, backgrounds} 
\usepackage{adjustbox}
\usepackage{epstopdf}
\usepackage{multirow}
\usepackage{wasysym}
\usepackage{gensymb}
\usepackage{amssymb}
\usepackage{pifont}

\usepackage{dblfloatfix}

\usepackage{amsthm}

\newcommand\copyrighttext{%
  \footnotesize \copyright{} 2016 IEEE. Personal use of this material is permitted. Permission from IEEE must be obtained for all other uses, in any current or future media, including reprinting/republishing this material for advertising or promotional purposes, creating new collective works, for resale or redistribution to servers or lists, or reuse of any copyrighted component of this work in other works. The official version can be found at \url{http://dx.doi.org/10.1109/GHTC.2016.7857261}}
  
\newcommand\copyrightnotice{%
\begin{tikzpicture}[remember picture,overlay]
\node[anchor=south,yshift=10pt] at (current page.south) {\fbox{\parbox{\dimexpr\textwidth-\fboxsep-\fboxrule\relax}{\copyrighttext}}};
\end{tikzpicture}%
}

\begin{document}
\title{Maintaining both availability and integrity of communications: challenges 
and guidelines for data security and privacy during disasters and crises}

\author{\IEEEauthorblockN{Flor \'{A}lvarez and Matthias Hollick}
\IEEEauthorblockA{Secure Mobile Networking Lab \\ TU Darmstadt, Germany\\
Email: \{falvarez, mhollick\}@seemoo.tu-darmstadt.de}
\and
\IEEEauthorblockN{Paul Gardner-Stephen}
\IEEEauthorblockA{School of Computer Science, Engineering \& Mathematics,\\
Flinders University, Australia\\
Email: paul.gardner-stephen@flinders.edu.au}
}
%\copyrightnotice

%Telephone: (800) 555--1212, Fax: (888) 555--1212}

\maketitle
\copyrightnotice
\begin{abstract}
%!TEX root = ../paper.tex
Communications play a vital role in the response to disasters and crises. 
However, existing communications infrastructure is often impaired, destroyed or overwhelmed during such events. 
This leads to the use of substitute communications solutions including analog two-way radio or unsecured internet access.
Often provided by unknown third parties, these solutions may have less
sophisticated security characteristics than is desirable. 
While substitute communications are often invaluable, care is required to minimize the risk to NGOs and individuals stemming from the use of communications channels with reduced or unknown security properties.
This is particularly true if private information is involved, including the location and disposition of individuals and first responders.
In this work we enumerate the principal risks and challenges that may arise, and provide practical guidelines for mitigating them during crises.
We take plausible threats from contemporary disaster and crisis events into account and discuss the security and privacy features of state-of-the-art communications mechanisms.

\end{abstract}

    % Load the content from subfolder
    %!TEX root = ../paper.tex
\section{Introduction}
\label{sec:introduction}
\subsection{Motivation}
Communications is an important ``force-multiplier'' during disasters
and crises \cite{comfort2006communication}. The challenge, of course,
is that during such events communications capability is typically reduced, 
while, conversely, demand for communications increases. This often creates, 
or broadens, the gap between what the indigenous communications capacity can
provide, and the demands placed upon it. As a result supplementary
communications capabilities are often bought into disaster and crisis
zones in an attempt to bridge this gap. A problem arises, however, in that the security properties of the
collection of communications options available in a disaster zone may
not be immediately apparent, and as adversaries become more and more
active during disaster response activities \cite{nzherald}, there is a compelling need
to provide responders with practical evidence-based information that
will allow them to make informed decisions about their use of the
available resources, so as to maximize their mission effectiveness,
while minimizing the risk for harm.
\subsection{Contribution}
The primary contribution of this paper is to translate understanding
generated through recent information security research into a form where it
can be of direct use by those engaged in providing and using
communications technologies during disasters and crises.
To address this goal, we analyze and identify the most significant 
security requirements, possible threats and challenges for
communications technologies used during disasters and crises.
We also summarize several existing technologies utilized by responders
as communication channels during disaster, their assumptions as well
as security features they offer. Finally, we provide, using a flow chart, 
a useful guideline to select a suitable technology, which 
takes into account possible attacks that could arise during disaster scenarios.
\subsection{Outline}
This paper is structured as follows: first, representative communication
scenarios in a disaster situation are presented. Next, adversary 
and threat models are discussed. Subsequently, the main security 
features considered in this paper are analyzed, followed by a description of the 
communication technologies used in emergency scenarios and their security 
weaknesses.
Finally, practical guidelines for mitigating the risks and challenges 
from the use of different communication channels during crises are discussed.
We conclude the paper providing potential future research directions.
\subsection{First of a series}
In recognition that this is a constantly evolving situation, and that it is impossible for any one 
paper to exhaustively explore this space, it is our intention that this guidelines paper is the first 
in a series. Successors will attempt to include further technologies and responses to the predominant 
prevailing threats and explore mitigation strategies as they evolve. 

	%!TEX root = ../paper.tex
\section{Scenarios}
\label{sec:scenarios}
There is naturally partial commonality between information security
scenarios between crisis and non-crisis contexts. The purpose of this
paper is, however, to address the particular needs of crisis
responders. In this section we define what we mean by crisis, and
enumerate scenarios that are known to occur in crisis situations.
\subsection{Definition of a disaster or crisis}
There is considerable variation in the definition of a disaster, both
in terms of the scope of the kind of events included, as well in the
nomenclature used \cite{mayner2015defining}. Authors in \cite{mayner2015defining} explored the
existing literature covering the definition of the term disaster,
discovering hundreds of variant definitions. From those, they
proposed the following as the most consistent definition of a 
disaster, a definition which shall be assumed throughout this paper to
cover the events of interest. In this paper, a crisis is considered to be a variant descriptor for a disaster.
\begin{displayquote}
`..the widespread disruption and damage to a community that exceeds
its ability to cope and overwhelms its resources.' \cite{mayner2015defining}
\end{displayquote}
This procedurally generated consensus definition is in strong
agreement with the definition produced by the United Nations for
disaster:
\begin{displayquote}
`..A serious disruption of the functioning of a
community or a society involving widespread
human, material, economic or environmental
losses and impacts, which exceeds the ability
of the affected community or society to cope
using its own resources.'\cite{UNISDR2009}
\end{displayquote}
That is, the overarching defining characteristic of a disaster, as compared to,
for example, an accident or other non-disastrous undesirable event, is
that in a disaster, there is a need for external support. In the case
where a disaster involves communications, this implies that additional
communications capability or capacity is required to be introduced to
the disaster zone. This may then require that responders, residents,
existing services and authorities, and any other parties active in the
disaster zone, resort to communications tools, technologies, suppliers and/or
media that they would not normally use. As a result, the resulting communications 
may have drastically different security properties compared with the regular use. 
It is from this context that the following scenarios
are derived.
\subsection{Use of unknown or insecure internet connectivity}
Perhaps the simplest scenario is using internet connectivity, possibly provided
by unknown parties, for whatever purposes are necessary.
Here the scenario is simply that the security properties of the
internet connectivity are unknown. It may also be that existing
internet connectivity is being used, but which is not adequately
secured. For example, disaster responders may come into a disaster
zone, and make use of existing indigenous internet connectivity, that
may not be as secure as they are normally accustomed to.
Possible risks include that the
connectivity may be surveilled, censored, inter-mediated or otherwise
interfered with by some party, whether or not they are the supplier
of the internet connectivity. For example, a militia or other informal
power-block may monitor internet communications in order to gain
advantage or further their victimization of others. In some locations
it is also possible that surveillance may be used to gather material
in order to seek expulsion or incite or commit extra-judicial violence or other
actions against disaster responders, e.g., in areas where religious or
other extremist activity is present.
\subsection{Communications between personnel within organizations}
Where conventional communications are disrupted, it may be necessary
to use personal communications media that are less secure than those
ordinarily used. For example, analog or digital radios may be used in
place of cellular mobile telephony. Or alternatively, mobile
telephony standards may be used that offer inferior security properties, such as 2G (and
some 3G) GSM networks. In such cases, many of the same risks can
arise as when using internet connectivity. That is, communications may
be eavesdropped, falsified, or otherwise interfered with.
Eavesdropping alone can pose a significant risk to personnel, when it
allows hostile actors to anticipate their movements and disposition.
A complication that can arise is that in some locations it may be
desirable for an organization to communicate in the clear, so as to
avoid causing anxiety for the government in the disaster zone.  
\subsection{Communications between organizations}
Communications between organizations is in many regards isometric to
communications within organization, but with any negative impacts
being able to affect more than a single organization.
\subsection{Dissemination of public safety or similar information}
Public dissemination of information differs strongly from the above
scenarios, where confidentiality is a prominent requirement. Instead,
for public dissemination of information it must be readable by all
parties, and thus sent in the clear, but it must also be authentic,
that is not impersonable, and it must not be able to be modified,
which could lead to dangerous mis-informing of recipients. A
further issue is timeliness of the information, so that only
information that is still correct and useful is received, and cannot
be replayed at a later time by a hostile party.

	%!TEX root = ../paper.tex
\section{Adversary and Threat Models}
\label{sec:adversary}
A user in a communication network during a disaster can act in any of the following 
roles: (1) an honest user, who contributes to the communication, cooperates with 
neighbors, acts according to the specified protocols, rules, etc, or, (2) a malicious user, 
who tries to manipulate or subvert the normal communications.
An adversary's behavior and the impact of an attack can vary according to the 
capabilities of the adversary as well as the scenario or situation in which an attack 
occurs. 
Whether an adversary works alone and independently (\textit{single attacker}), as 
compared to whether they collaborate with other malicious users (\textit{colluding attacker}), 
this will depend on the goals of an attack, but also on the motivation of these attacks. 
On the one hand colluding attackers can be desperate people, who try to take 
advantage of available resources without regard for others, for example, to meet 
their basic needs for food, shelter and other materials. For these actors, they can 
perhaps be modeled from a game theory perspective as seeing the situation as a 
zero-sum game, and they are seeking to maximize their gain, without being arrested 
by the fact that it necessarily increases the loss of other parties. That is, their objective
is their gain, rather than the loss of others. On the other hand, there can be actors such 
as terrorists, who can also be colluding attackers, but who are actively trying to exploit 
destruction and chaos in the civilian communication causing panic and confusion, and can 
perhaps be modeled from a game-theory perspective as seeking to minimize the sum of 
the game from the perspective of other parties. That is, rather than failing to be arrested 
by the presence of a zero-sum game, they are actively motivated by this. As this analysis 
reveals, while their motivations, and thus modes of operation may differ, they have the 
same final effect of undertaking actions that induces losses upon other parties.
In this subsection we are going to give a high level overview of possible adversaries that can be considered during a disaster scenario:
\begin{itemize}
\item \textit{Non-cooperative user's behavior: } In some situations when normal users (civilian or organization) compete to gain some limited resources, they may cease cooperative actions, as they actively seek to optimize their own resources. Their behavior can affect the entire communications channel as a whole.
\item \textit{Militias: } Haiti Earthquake of 2011 saw many examples of this kind of adversary \cite{nelson2010media}. Militias broke into stores and induced a state of violence and anarchy. Without a security mechanism for communication channels, the militias are able to listen to the communications of others, and use this intelligence to their advantage, and therefore the disadvantage of others \cite{yates2011emergency}. 
\item \textit{Terrorists: } A terrorist has different purpose that vary according to the situation: creating fear in the population, gaining international publicity for a terror group, but also supporting political or religious ideology, perceived or actual inequitable treatment, among other factors \cite{szocik2016axiological,isaac2015economic}.
\item \textit{Looter: } During the Katrina Hurricane of 2005, fabricated reports of the shooting of rescuers and civilians, act of violence, etc., were reported \cite{latimes2005}. The spread of rumors about looting and the lack of authority's presence contributed to a general rise in panic and wide distribution of inaccurate information to citizens. Attackers used this information and looted abandoned properties or tried to take the resources of isolated community residents.
\item \textit{Political motivated organizations: } In the Russian - Georgia conflict of 2008, governmental and civilian infrastructure was the victim of cyber attacks, whose main purposes was to disrupt and to compromise the communication  within Georgia, as well as to gather intelligence from and about military and political groups \cite{hollis2011cyberwar}. 
\end{itemize}
In this section, we describe possible mechanisms that can be used by an adversary in order to achieve an attack:
\begin{enumerate}
\item \textbf{Eavesdropping: } Passive eavesdroppers will attempt to intercept the location information or movement patterns of users, e.g., to collect social graphs of participants. A malicious user can use this information for subsequent attacks. 
\item \textbf{Message tampering: } Attackers can inject false location information or modify legitimate information affecting the reliability or truth in such networks. For example, an adversary may distribute fabricated information about available local infrastructure (e.g., hospitals or shelters locations) to overload them.
\item \textbf{Identity theft: } Adversaries may pretend to be valid users, or other legitimate identities. For example, they could pretend to be a trusted authority, such as the government or a well-known NGO, and send false information about a disaster which does not exist, causing panic among the populace.
\item \textbf{Jamming: } Attackers may block future packets or the communication channels between normal users. 
By deliberately interfering in this way, users can be prevented from communicating with other devices,
 and produce a lack of availability. Single or colluding attackers may try to interfere with data sent
 by authorities, e.g., warnings or announcements about potential hazards, to further amplify the effect
 of their attack. For example, by maintaining the impression that the communications channels are operative, e.g., by allowing all other communications to continue unimpaired, thus reinforcing the implicit belief among other parties that there are no suppressed communications.
\end{enumerate}

	%!TEX root = ../paper.tex
\section{Information Security Model}
\label{sec:securitymodel}
We identify the main security properties considered in this paper when choosing a specific system for a crisis. 
\subsection{Confidentiality (that no one else can listen in)}
There are scenarios, where it is important to ensure that only the specified person can read the exchanged information \cite{nelson2010media, hollis2011cyberwar}. Attacks to this property attempt to get access to sensitive information without accurate authorization. Hence, different methods and mechanisms are necessary to keep the content of a message secret from unauthorized users. 
In certain scenarios, however, where the data is public by itself, this property may be not mandatory, or even desirable.
\subsection{Integrity (that communications cannot be modified)}
Attacks to the integrity of communications attempt to modify legitimate information, for example replacing the text of a communications with text of the attackers choosing. In this context it is necessary to verify if the exchanged information between two parties was altered during transmission, as well as to protect the content of a message against any alteration. 
In most cases, this is a mandatory property, as otherwise an attacker could maliciously cause arbitrary mis-information.
\subsection{Authenticity (that no one can impersonate another)}
The receiver of a message should be able to corroborate that the purported sender is in fact the authentic author of the communication \cite{latimes2005}, e.g., for public alerts or warnings from rescuers. Without such an authentication mechanism, users could impersonate one another, with considerable scope for disseminating mis-information with malicious intent.
\subsection{Validity (that a communication is fresh, and has not been superseded by another)}
In some scenarios it is important to verify if the exchanged data is still valid data \cite{latimes2005}. Without a mechanism to verify the freshness of the information, 
an adversary can replay old messages without being detected, and the old data may not merely be of reduced value, but may in fact be harmful. For example, an attacker could re-play messages indicating 
that particular places are safe (or unsafe) to influence the activities of people and/or 
agencies to their advantage.
\subsection{Anonymity, Privacy, Social Accountability and Non-Repudiability}
Anonymity, privacy, social accountability and non-repudiability represent
an additional set of security properties that are particularly
important when private persons communicate with one another, with the
public at large, and in some cases, with government and authorities.
Together, they allow users to interact without excessive fear of the
consequences of their communications, or of the consequences of the
communications activities of others (including passive collection of
communications).  However, the focus of this paper is on
communications among persons and organizations where the opposite is
the case: where it is highly desirable for the authenticity of persons and
their actions to be sustained.  Therefore these topics are not within
the scope of this paper.
However, they are not necessarily irrelevant to all disaster scenarios: their importance can differ according to the nature of the emergency scenario. 
For example, anonymity can be of significant importance in scenarios where a government controls communication channels. In such cases, it is important that the opposition or other people with differing opinions or anti-government positions cannot be identified, or at least cannot be distinguished among a group of senders.
An example of this is following the 2008 Sichuan Earthquake in China, where a teacher was sentenced to one year of prison, because he posted photos about the damage occasioned by the earthquake on schools \cite{sichuan2008}.

	%!TEX root = ../paper.tex
\section{Communication technologies}
\label{sec:commtech}
Different technologies can be used in the communication between organizations 
and civilians after disaster scenarios. These systems, however, may suffer a lack of security,
providing either none or weakened security feature(s). This is particularly the case
for security features that depend on some kind of centralized infrastructure.
For example, many communications systems offer security 
features that depend on the availability of centralized infrastructure, e.g., internet 
access is often required to check the validity of digital certificates. 
The following text describes a number of the communications technologies that are often
provided post-disaster, and explores their security properties.
\subsection{Infrastructure-less technologies}
\subsubsection{Analog Radio}
Analog radio has played a prominent role in disaster communications
for many decades, and as our private correspondence with New Zealand
Red Cross confirms, this role continues to the present. In particular,
\gls{uhf} and \gls{vhf} hand-held radios are frequently used to provide
communications among personnel deployed during and following
disasters. \gls{hf} radio is still present in many situations, however NZ
Red Cross informs us that its role is diminishing, and is increasingly
being replaced in the field by lower-cost satellite based-solutions,
such as the deLorme inReach \cite{inreach}, that allows global
communications reach, without the complications of maintaining and
operating an \gls{hf} radio installation, including ensuring that trained
personnel are available.

Analog radio is also an unsecured broadcast medium. That is, any
party with a suitable radio is able to listen to all
communications. Indeed, all parties must listen to all communications
if they are to hear communications that are intended for them.  
In some situations this broadcast nature is helpful, for example, when
information is non-confidential, and desirable for multiple parties to
receive, e.g., the disposition of members of a team as they carry out
their activities.  However, even in such cases, 
adversaries are able to easily listen to communications, replay recorded
communications (simply by recording the transmissions of others on a
channel, and then replaying them into the microphone of a radio at a
later time), or to more actively participate in communications in
a variety of subversive, or even merely disruptive manners.  This is
because analog radio is lacks confidentiality, authenticity and integrity.

\subsubsection{Digital Radio}
The arch-typical example of this technology in Europe is \gls{tetra}, 
which allows the exchange of speech and status messaging with a limited data rate. 
Digital radio systems offer some improvements over analog radio, in
that communications often supports confidentiality, and
some digital radio systems can authenticate communications, and ensure
their integrity. However, this is not the case for all digital radio
systems, and care must be taken to gain an adequate understanding of
how confidentiality, authentication and integrity are provided by the
radio system, and how they can be protected against common attack
paths, such as theft, loss of a radio handset or traffic analysis, which
imply a significant threat in some environments, e.g., in presence of militias. 
Also, some digital radio systems require supporting infrastructure, such as a base station or central repeater.

\subsubsection{Off-grid ad-hoc networks} 
In addition to the traditional analog and digital radio communications
system, distributed \gls{manet} have been 
developed over the past twenty years or so, 
primarily enabled by the development of 802.11 Wi-Fi 
\cite{giordano2002mobile,kannammal2016survey}.  
The general intent of such networks can be summarized as 
facilitating the creation of networks, without reliance on
conventional fixed infrastructure.  
Within the humanitarian space, projects of potential interest include, without limitation, the
FreiFunk and associated projects in Europe \cite{freifunk,hardes2015performance}, the Commotion
Wireless project from the USA \cite{commotion,namiot2015mobile}, and the Serval Project from Australia 
\cite{gardner2011serval,gardner2012meshms,gardner2013rational,gardner2013serval}.

Each of these projects has a particular focus, for example, FreiFunk and the related 
movement can be perhaps simply summarized as facilitating the
provision of (typically wireless) internet access, independent of
existing infrastructure. The Commotion Wireless project is primarily
interested in providing secured internet and intranet access in
difficult environments, such as providing communications for
dissidents under oppressive regimes. The Serval Project is primarily
focused on providing secured mobile telecommunications, without
dependence on existing infrastructure, with an emphasis on
disaster response and isolated communities.  
The differing approaches and intentions of each projects results in a
diversity of security properties of these networks. However, many
modern incarnations pay particular attention to the issues of
confidentiality, authenticity and integrity of communications---including the difficult matter of managing the routing of
communications among the devices in a \gls{manet} \cite{reina2014survey, di2014security}. 
When contemplating using such systems, particular emphasis should be placed on 
ensuring an appropriate fit for the security requirements at hand.

\subsection{Infrastructure-based technologies}
\subsubsection{Cellular Networks}
Cellular networks were designed to provide voice applications based on
digital technology, and offer improved confidentiality, authentication
and integrity compared with the first generation analog cellular
networks. While the situation has improved
with each successive generation of cellular technology, significant
security issues remain and are constantly uncovered. 
Different studies and important examples \cite{ricciato2010review, 
ja2006analysis, enck2005exploiting} have 
demonstrated the feasibility of some attacks exploiting successfully
the existing vulnerabilities and weaknesses of the underlying standards or 
poor operational practice by operators, e.g., by redirecting user traffic or 
theft of valid user identities, particularly in the face of a determined
adversary who has the means to obtain specialized hardware that allows
interception of 2G and 3G communications with relative ease. Such
hardware is available for less than US\$1,000, placing such attacks
well within the reach of many adversaries.

Also, cellular networks are
designed with ``lawful intercept'' capability included, and therefore
any use of such networks must take into account that not only the
local government, but also potential adversaries with sympathizers
within cellular carriers, may be able to intercept all communications,
and use the advanced location capabilities of modern cellular
base-stations to obtain pervasive location data on users, including
reliable predictions of where users are likely to be at a future time
and date. This can be particularly hazardous in civil unrest
situations, where one or more of the belligerents have reason to be
opposed to the delivery of humanitarian relief in an area. For
example, the authors are aware through private correspondence of
situations where the resources of cellular networks and/or ISPs are
alleged to be made available to such parties, and where that information 
has apparently been used to target relief workers.

\subsubsection{Satellite Radio}
Satellite communication can offer high-speed data and video
transmission in crisis. Many systems, however, suffer problems such as weak 
security properties, problems with synchronization, and 
in some countries, their use during disasters may not be allowed. 
Notwithstanding the above difficulties, newer satellite communications
tools, such as \gls{sbd} modules connected to the
Iridium constellation offer the ability to communicate from most
locations on the surface of the earth using an SMS-like interface, and
only a small battery powered satellite terminal that can be carried in
a pocket, and can pair with a smart-phone. 
However, the security on the Iridium \gls{sbd} service has to be described by 
`security by obscurity'.

There is sound reason to believe why a determined attacker 
could listen to all Iridium \gls{sbd} data directed to a given locale, by 
demodulating the broadcast signal from each satellite as it passes overhead, 
using only a few hundred dollars of equipment \cite{iridiumsbdhack}.  
This attack is particularly concerning, because it would seem that the SMS-like
service is unencrypted, and could be relatively easily spoofed. That
is, a determined adversary could potentially transmit signals that an
Iridium \gls{sbd} would interpret as having come from a satellite, thus
allowing an adversary to inject themselves arbitrarily into
conversations. Because the transmissions from the satellites are
unencrypted, this could be implemented in a manner that uses
intelligence gathered from the transmissions to allow the mounting of
sophisticated attacks that would be difficult for an end-user to detect.

\subsubsection{Social media}
In recent years, different disaster scenarios have widely shown 
the importance and benefits of social media (e.g., twitter, facebook, 
webblogs, wikis and other web-based resources) in such situations 
\cite{nelson2010media, yates2011emergency, peary2012utilization, 
toriumi2013}. 
These media enable a communication channel between the different 
actors in the disaster area, but also facilitate the coordination, 
organization and involve spontaneous volunteers and civilians in 
the rescue efforts.  
Platforms such as crowd-sourcing and related mapping tools have begun to play 
prominent roles following disasters, allowing important information to be collected, 
classified and organized, greatly increasing its value and utility in the response effort.

These communications channels, however, build on existing communications 
infrastructures---mainly the internet---with a central infrastructure, which itself may be affected after a disaster.  
Due to the huge volumes of data and redundant information generated by such media, 
the identification of relevant messages without an appropriate prioritization of the information
can be difficult, affecting the ability for a prompt and efficient response by rescuers organizations. 
An interesting example was Hurricane Sandy in 2012, 
where more than 20 million tweets about the disaster were generated \cite{sandy2012}.
Furthermore, the open nature of social media (every person 
can publish new information, or repeat existing information) complicates the validation 
of the data being spread, making it easier for malicious users to manipulate data, or in 
some cases, cause the spread of false information. The Safety Check 
feature activated by Facebook after the 
Nepal earthquake in 2015, was a good example of the misuse of social media during 
disaster scenarios \cite{NewsSC2015}.
\begin{table*} [!htb]
\caption {Summary of communication technologies and their security features in disaster scenarios}
\label{tab:tech}
\centering

	{\scriptsize{
\begin{tabular}{p{2cm} p{1cm}
						p{1cm}ccp{9cm}}
\toprule 
\multicolumn{2}{c} {\textbf{Technology}}& 
\multicolumn{1}{c} {\textbf{Examples}}& 
\multicolumn{2}{c} {\textbf{Security Features}} & 
\multicolumn{1}{c} {\textbf{Comments}} \\ 

\multicolumn{2}{c}{} & &
{\textit{voice}} & 
{\textit{text}} & 
 \\ 
 \rowcolor[gray]{.8} 
\multicolumn{6}{l}{\textbf{\textit{Infrastructure-less Technologies}}} \\ 
\multicolumn{2}{l}{\raisebox{-5ex}{Analog Radio}} &  
                \gls{hf} radio    & 
             0 & 
              n.a. & 
              In situations where it is desirable to send non-confidential information and to reach 
              multiple parties, the broadcast property of analog radio is very helpful. Due to the 
              broadcast nature of this medium, however, any party with a suitable radio is able to 
              listen all communications, or replay old recorded communications.
              That is, confidentiality, integrity, authenticity and validity are not provided by analog radio.    \\ 
\hline 
\multicolumn{2}{l}{\raisebox{-5ex}{Digital Radio}} & 
			TETRA, Project 25
                    & 
           C*, I*, A*,
           & 
           C*, I*, A* & 
              Digital radio provides several improvements over analog radio. Some level of security 
              such as confidentiality, authentication, etc, can be provided by these radio systems. Nevertheless, different  
              analyses \cite{clark2011, glass2011} have identified significant security weaknesses 
              of these system, which lead to potential attacks that could affect the provided security features 
              such as confidentiality, etc.
              For example, an adversary can get access to sensitive data, track devices or analysis exchanged traffic. 
              Furthermore some digital radio systems still require some form of supporting central infrastructure.\\ 
\hline 
\multicolumn{2}{l}{\raisebox{-5ex}{Off-grid ad-hoc networks}} & 
            Commotion Wireless, FreiFunk, Serval Project        & 
            C*, I*, A* & 
            C*, I*, A* & 
            \glspl{manet} have emerged over the past twenty years to enable a communication without reliance on 
            traditional fixed infrastructure. Different projects focused on \glspl{manet} as a viable solution to implement 
            multi-hop communication in emergency scenarios \cite{reina2014survey, di2014security}.  These
            technologies have the potential to provide advanced security properties, e.g., confidentiality and authentication, 
            without recourse to any centralized or fixed infrastructure \cite{bang2013manet},
            however care should be taken to verify that the desired properties are provided by the particular system being considered. \\ 
\rowcolor[gray]{.8} 
\multicolumn{6}{l}{\textbf{\textit{Infrastructure-based Technologies}}} \\
\multicolumn{2}{l}{\raisebox{-5ex}{Satellite Radio}} & 
           Iridium, VSAT, Inmarsat-C  & 
           C*, I*, A* & 
           C*, I*, A* & 
			Satellite communications are widely used in different scenarios, including in emergency communications. 
			In 2014 IOActive \cite{ioactive2014} evaluated several satellite communication systems 
			and identified numerous vulnerabilities that attackers could exploit. For example, attackers can disrupt all 
			communications on some satellite systems by compromising just a few devices. \\
\hline
\multirow {3}{*}{ } &
     					 &
                          2G & 
                    0 & 
                    0 & 
                    2G Cellular networks suffer from inherent security weaknesses. These weaknesses
                    arose during the rapidly growth of cellular networks, particularly as they began to carry data as well as voice. Existing vulnerabilities such as affecting the availability, 
                    redirecting user traffic or cloning valid user identities, represent several of the readily exploitable security flaws 
                    of 2G networks. Although 2G networks use different encryption methods, some of the mandatory mechanisms do not provide 
                    any security and are vulnerable to some attacks. 
                    Sound operational security is of utmost importance in 2G networks.\\ 
                     \cline{3-6}
                     
                     \raisebox{-5ex}{ \textit{Cellular Networks}}    & 
                      &
                        3G  & 
                     C*, I*, A* & 
                    C*, I*, A* & 
	                3G Cellular networks provide improvements, new functionality and security features over 
	                2G networks. These networks, however, also inherit some vulnerabilities and threats from 2G,
	         		 e.g., flooding attacks, which exploit paging vulnerabilities in both cellular network generations. 
	         		 3G cellular networks implement different countermeasures to avoid many of the 2G 
	         		 vulnerabilities. These countermeasures, however, have no effect in dual operation mode, that is
	         		 when 2G and 3G networks employ the base stations simultaneously (inter-operation 
	         		 between 2G/3G networks), which is still common in many countries \cite{ricciato2010review}. \\ 
                     \cline{3-6}
                     
                     & 
                     &
                        4G  & 
                     C*, I*, A* & 
                    C*, I*, A* & 
                    The security measures of 4G networks are very similar to those of the 3G Cellular networks. The main additional 
                    security improvements of 4G networks are focused on key management and protection against 
                    physical attacks against base stations. Nonetheless, some security issues still remain a problem in the 4G networks. 
                    Some studies have already identified security vulnerabilities in the architecture of 4G networks \cite{park2007survey, bikos2013, cao2014}.
                     \\ 
                     \cline{3-6}
\hline 
\multirow {3}{*}{ } 
					& Online Social Networks &
                       Twitter, Facebook   & 
                       0 & 
                       0 & 
                    The use of social media such as twitter or facebook in disaster scenarios, has increased rapidly. 
                    Such networks provide many opportunities to users e.g., to share information, photos or news, 
                    and to organize each other. Their wide availability and diverse facilities make them valuable to spread 
                    information widely in a short period of time. However, those same characteristics can be abused by adversaries,
                    e.g., by creating multiple fabricated identities and using those to spread false information \cite{zhang2010}.
                     \\ 
                     \cline{2-6}
                     
					\raisebox{-3ex}{\textit{Social media    }} 
                     & Email service &
                         Gmail, Yahoo & 
                       n.a. & 
                    C*, I*, A* & 
                     Communications media that can operate with only intermittent
                     connectivity, such as email, are also important tools in disaster response, allowing some form for 
                     data exchange between different parties. Nonetheless, the internet based platforms are provisioned and 
                     operated by untrusted third parties, and their security depends on the service providers
                     \cite{gao2011}, thus care should be taken to select appropriate providers.  As a minimum email providers
                     should be selected that always use HTTPS to protect the privacy of all communications..\\
                     \cline{2-6}
                     
                    &  
                    Chat, VoIP systems &
                        Skype, Whatsup  & 
                      C*, I*, A* & 
                      C*, I*, A* & 
					   Alternative systems such as VoIP or chat applications are very helpful during emergency scenarios. They do 
					   not only offer the same features from traditional communication channels, e.g., PSTN, but can also provide additional 
					   capabilities, e.g., video and large group calling. Nevertheless, many of these implement proprietary protocols and 
					   security countermeasures, which imply interactions between different technologies that can also lead to 
					   unexpected results which may in turn result in security issues \cite{keromytis2011, keromytis2012}. \\
\bottomrule
\end{tabular} 
\begin{tabular}{c}
      \raisebox{-2ex}{ \textbf{Legend: } \textit{0 = considered insecure};  
															    \textit{n.a. = not applicable}; 
															    \textit{C = Confidentiality of data}; 
															    \textit{I = Integrity of data}; 
															    \textit{A = Authenticity of sender}; 
															    \textit{* = known issues}}\\ 
    \end{tabular}
	}}  
\end{table*}
%end section communication technologies

	%!TEX root = ../paper.tex

\section{Guidelines}
\label{sec:guidelines}
In this section, we summarize the main communication technologies utilized during 
disaster scenarios and known security threats found in the literature.

Table \ref{tab:tech} compares communication technologies discussed in Section 
\ref{sec:commtech}. It provides selected references to relevant information security surveys or known 
vulnerabilities, as well as security features offered by 
each technology (depending on configuration). We were only able to find relatively few studies or related work that supply directly helpful 
information about 
possible issues, and security requirements for such technologies in disaster scenarios. 
The summarized technologies are not necessarily applicable to all 
disaster scenarios, their importance and requirements can differ according to the features of the 
emergency scenario, as well as by the communication mode(s) required. 
Overall, the results in Table \ref{tab:tech} indicate the lack of security for several 
communication technologies, e.g., on analog radio communication, 2G cellular networks, etc., 
which nonetheless continue to be major communications channels used during disasters.

Due to the wide range of scenarios and their specific characteristics, it is difficult to generalize and mandate 
specific technology choices and security settings. Likewise, it is impractical to specify requirements for all 
imaginable disaster scenarios in detail.  We simply encourage practitioners to educate themselves of these issues,
so that they can make informed decisions relevant for their situation.
\begin{figure}[h]
\qquad
	\caption{Flow chart for the selection of appropriate communications technologies for use during disaster scenarios}
	 \label{fig:guidelines_flow}
   	\centering {
\vspace{1em}
      	\includegraphics[width=.5\textwidth]{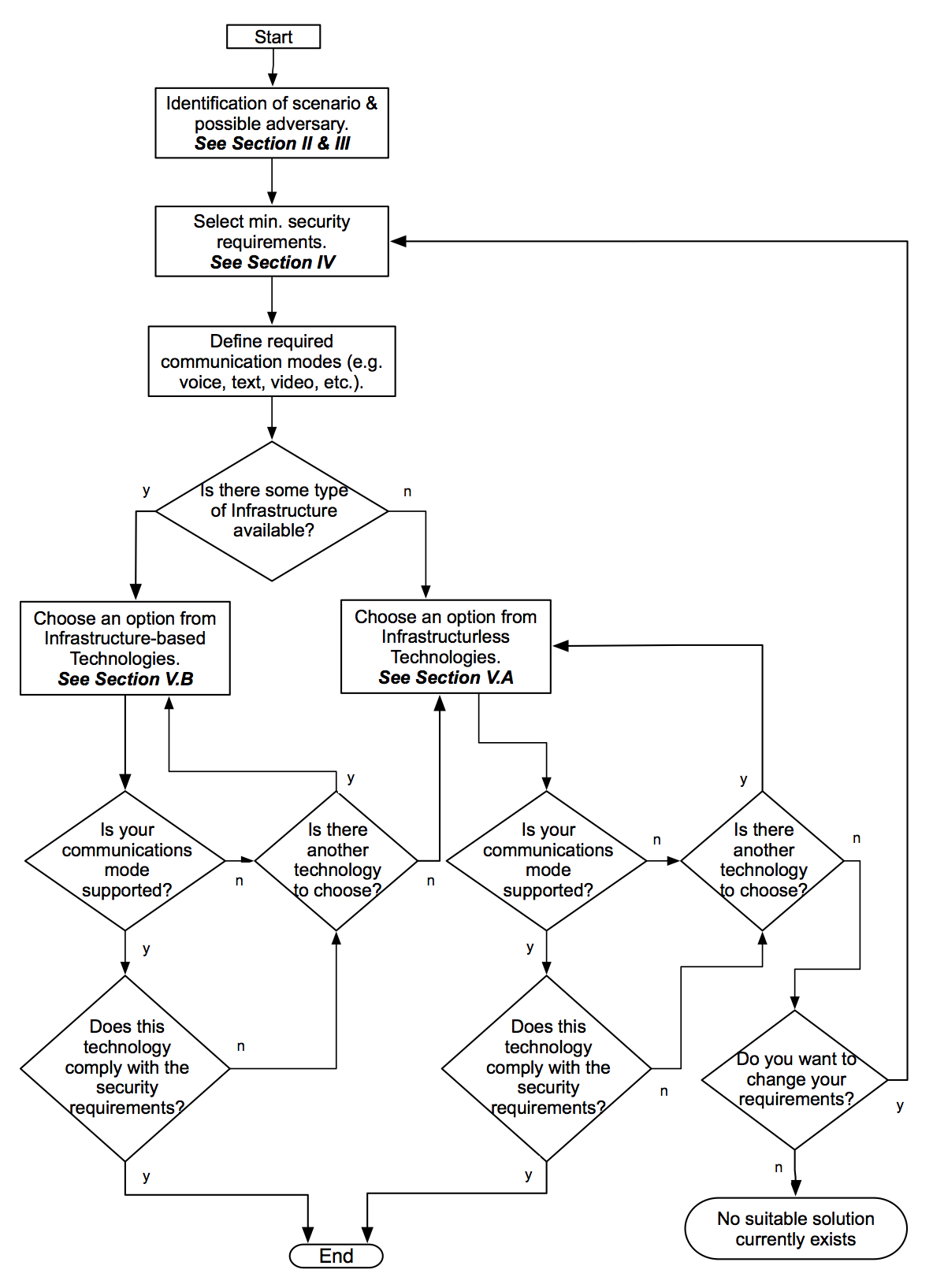}}
      	\qquad
\end{figure}

In support of this, we have attempted to identify the minimal steps necessary to rationally select an appropriate technology 
during crisis (Fig. \ref{fig:guidelines_flow}).

First of all, it is important to identify possible security threats that can arise in the crisis. 
Then, one can define which communication modes are required (e.g., to transmit voice, etc). 
Following this, one can also consider and specify the minimal security requirements that a communication 
technology should provide in order to minimize the impact of possible adversaries identified in the 
first phase. 
Finally, one can select a technology or technologies from all available technologies and 
verify if the technology complies with the identified requirements.

% section guidelines (end)

	%!TEX root = ../paper.tex

\section{Conclusion}
\label{sec:conclusion}

Communications technologies for emergency response predominantly focus 
on the establishment of communications after a disaster. 
Security is often completely ignored due to missing knowledge on security 
mechanisms and the complexity to implement secure communications.
Nevertheless, foundational security features, like authentication, 
confidentiality and integrity become mandatory, as without those features significant risks result, e.g. 
false information can be distributed during crisis situations, resulting in confusion or 
a loss of trust amongst recipients. 

To address these issues, this paper has presented an overview of some 
scenarios that occur in crisis situations, where crisis responders use insecure 
channels. Furthermore, we have provided an analysis of possible threats that may 
appear during a disaster, defining an adversary model. Hence, we have described the 
most significant security requirements to achieve a practicable and secure 
communication after a disaster scenario.

Finally, we propose a flow chart (see Figure \ref{fig:guidelines_flow}) which 
summarizes the different steps considered necessary 
for choosing an appropriate technology to communicate during a disaster (see technology overview in Table \ref{tab:tech}). 
It is our hope that these contributions may be of practical assistance to emergency and disaster responders.

\section*{Acknowledgment}

% Rationalise and reorder as required
This work has been funded by the German Federal Ministry for Education and Research (BMBF)
within the SMARTER project. The authors would like to acknowledge the support of USAID (USA), the NLnet 
Foundation (Netherlands), Radio Free Asia (USA), the Shuttleworth
Foundation (South Africa), and the
Australian Foreign Aid Program within the Australian Department of
Foreign Affairs and Trade (Australia).



\bibliographystyle{IEEEtran}
\bibliography{References}

\end{document}